\documentclass[aps,rmp,onecolumn,
			   groupedaddress,superscriptaddress,
			   amsfonts,amssymb,amsmath,
			   citeautoscript,bibnotes,
			   a4paper]{revtex4}

\usepackage[pdftex]{hyperref}
\hypersetup{colorlinks,
	linkcolor={blue!75!black!80!yellow},
	citecolor={blue!75!black!80!yellow},
	urlcolor={blue!75!black!80!yellow},
	pdfstartview=FitH}
\usepackage[top=1in, bottom=1.25in, left=1.25in, right=1.25in]{geometry} 
\usepackage{graphicx}
\usepackage{amsmath}
\usepackage{amssymb}
\usepackage{mathrsfs}
\usepackage{xspace}
\usepackage{braket}
\usepackage{xr}
\usepackage{xcite}
\usepackage{xcolor,soul}
\usepackage{stmaryrd} 
\usepackage[UKenglish]{babel}
\usepackage{enumitem} 
\usepackage{diagbox}
\usepackage{float}
\usepackage{physics}
\usepackage{empheq} 
\usepackage{extarrows}
\usepackage{adjustbox}
\usepackage{multirow}
\usepackage[figuresright]{rotating}
\usepackage{rotating}
\usepackage{booktabs}
\usepackage{tabularx}
\usepackage{threeparttable}
\usepackage{adjustbox}
\usepackage{siunitx}
\usepackage{mathtools}
\usepackage{txfonts}  
\usepackage{txfontsb} 
\DeclareSIUnit[number-unit-product = ]\percent{\char`\%} 

\makeatletter
\def\frontmatter@authorformat{%
	\preprintsty@sw{\vskip0.5pc\relax}{}%
	\@tempskipa\@flushglue
	\@flushglue\z@ plus50\p@\relax
	\raggedright\advance\leftskip.25in\relax
	\@flushglue\@tempskipa
	\parskip\z@skip
}%
\def\frontmatter@affiliationfont{
	\small\slshape\selectfont\baselineskip10.5\p@\relax
	\@tempskipa\@flushglue
	\@flushglue\z@ plus50\p@\relax
	\raggedright\advance\leftskip.25in\relax
	\@flushglue\@tempskipa
}
\def\paragraph{%
	\@startsection
	{paragraph}%
	{4}%
	{\parindent}%
	{\z@}%
	{-1em}%
	{\normalfont\small\itshape\textsf}
}%
\renewcommand*\email[1][]{\begingroup\sanitize@url\@email{#1}} 
\makeatother

\setcitestyle{numbers,square,sort&compress} 

\makeatletter
\renewcommand*{\fnum@figure}{{\normalfont\bfseries \figurename~\thefigure}.}
\makeatother

\makeatletter
    \DeclareRobustCommand*{\deactivateaddvspace}{\let\addvspace\@gobble} 
    \DeclareRobustCommand*{\deactivatetocsubsections}{
    \def\l@subsection##1##2{}    
    \def\l@subsubsection##1##2{} 
    }
\makeatother

\makeatletter
\newcommand*{\addFileDependency}[1]{
  \typeout{(#1)}
  \@addtofilelist{#1}
  \IfFileExists{#1}{}{\typeout{No file #1.}}
}
\makeatother

\newcommand{\iu}{\mathrm{i}}

\newcommand{\appropto}{\mathrel{\vcenter{
			\offinterlineskip\halign{\hfil$##$\cr
				\propto\cr\noalign{\kern2pt}\sim\cr\noalign{\kern-2pt}}}}}

\newcommand{\ie}{i.e.\@\xspace}  

\usepackage{mathtools}

\makeatletter
\let\oldabs\abs
\def\abs{\@ifstar{\oldabs}{\oldabs*}}
\let\oldnorm\norm
\def\norm{\@ifstar{\oldnorm}{\oldnorm*}}
\makeatother

\usepackage{xifthen}
\usepackage{etoolbox}
\usepackage[normalem]{ulem}
\newboolean{togglecomments}
\newboolean{togglechanges} 

\setboolean{togglecomments}{true}
\setboolean{togglechanges}{false}

\newcommand{\comment}[2]{%
    \ifbool{togglecomments}%
    {\textcolor{blue!70!black}{\small\textsf{%
    \textsuperscript{\textsc{\textsf{\MakeLowercase{#1}}}}%
    [#2]}}} 
    {}}     
\newcommand{\swap}[2]{\ifbool{togglechanges}
    {#2}  
    {\textcolor{red!70!black}{[#1]}\textrightarrow{}\textcolor{green!50!black}{[#2]}}}
\newcommand{\remove}[1]{\ifbool{togglechanges}
    {}    
    {\textcolor{red!70!black}{#1}}}
\newcommand{\inset}[1]{\ifbool{togglechanges}
    {#1}  
    {\textcolor{green!50!black}{#1}}}
\newcommand{\optional}[1]{\ifbool{togglechanges}
    {}    
    {\textcolor{yellow!50!orange!80!gray}{#1}}}

\newcommand{\citeremind}[1]{%
    [\textcolor{blue!75!black!80!yellow}{
        $\blacksquare$%
	    \ifthenelse{\isempty{#1}}
	        {}
	        {\textsuperscript{\tiny\textsf{#1}}}%
	}]\xspace}

\newcommand{\usstaffil}{\footnotesize College of Optical-Electrical Information and Computer Engineering,\\
University of Shanghai for Science and Technology, Shanghai 200093, China}

\begin{document}

\title{Generation and high-resolution imaging of higher-order polarization via metasurface}

\author{Xiang Yuan}
\affiliation{\usstaffil}
\author{Hanming Guo}
\affiliation{\usstaffil}
\author{Songlin Zhuang}
\affiliation{\usstaffil}
\author{Jinbing Hu}
\email{hujinbing@usst.edu.cn}
\affiliation{\usstaffil}

\begin{abstract}
The generation and focusing properties of higher-order polarized beams have attracted  lots of interests due to its significant applications.
In this paper,we derived the formula of transforming linear polarization into higher-order polarization, which is applicable to generating arbitrary order polarization.
Based on the derived formula, the focusing properties of higher-order polarization by dielectric metasurface lens are studied , which exhibit an Abbe-limit-breaking feature for small numerical aperture, \ie, NA$<$0.6. 
When a binary phase (0 $\&~\pi$) is further imposed on the aperture of metasurface lens, the focusing spot of fourth-order polarization breaks Abbe limit even by 14.3$\%$ at NA$=0.6$.
In addition, the effect of fabrication tolerance, say, substrate thickness and central deviation, on the focusing  feature of higher-order polarization is also investigated.
Our study may find significant applications in achieving higher-resolution lithography and imaging, say, by just replacing conventional linear or circular polarization with higher-order polarization.
\end{abstract}

\maketitle

\setlength{\parindent}{0em}
\setlength{\parskip}{.5em}

\section{Introduction}
Vector beam is featured with vectorial phase distribution and spatially variant polarization~\cite{zhan2009cylindrical}, hence, compared to uniformly polarized beams such as linear and circular polarization, offers more degrees of freedom to interact with matter, which makes itself more valuable in practical applications, say, in particle trapping~\cite{ng2010theory}, optical communications~\cite{d2012complete}, and high-resolution lithography~\cite{hao2010phase}, optical information encoding~\cite{kumar2021arbitrary}.
Recently, higher-order polarized beams, a class of vector beams, have attracted increasingly attention~\cite{wang2021polarization,ruchi2020phase}.
Different from the class of vector beams featured with only phase vortex, higher-order polarized beams present the unique topological properties within the wavefront, such as, the Mobius strip~\cite{bauer2015observation,bauer2016optical}, the trefoil knot~\cite{larocque2018reconstructing,dennis2010isolated}, and optical skyrmions~\cite{tsesses2018optical,du2019deep}.
The diverse polarization distribution provides a versatile platform to explore novel optical effects and phenomena, facilitating the promising design of photonic devices~\cite{wang2021polarization}.

While there are some experimental reports on the generation of higher-order polarization, these methods are either complicated or limited in practical scenario.
For instance, Hasman $\mathit{et~al}$~\cite{niv2003formation} reported the generation of higher-order polarization from the incidence of circular polarization by using space-variant subwavelength grating.
However, this approach is only applicable to generate higher-order polarization of infrared range due to the fact that in order to realize the polarization conversion the maximum grating period must be much smaller than the desired light wavelength~\cite{niv2003formation}. 
The higher-order polarization could also be produced based on laser resonators and multi-beam interference, but this method is only applicable for lower-order polarization as the setup for higher-order polarization is complex~\cite{wang2021polarization,li2014efficient,pal2017generation}. 
One of the significant applications of higher-order poalrization is high-resolution imaging; Xu $\mathit{et~ al}$~\cite{xu2022tight} reported that higher-order polarized beam can focus into a circular spot with high resolution when the polarization order $p$ equals the topological charger $m$.
However, only analytical results were given therein; in the tight focusing case, there may be non-negligible difference between analytical and practical results. 
Therefore, an approach of compact setup that is experimentally simple and is applicable to generating higher-order polarization of arbitrary-order remains elusive.

In the present paper, based on half-wave-element metasurface, we derive the explicit expression of transforming linear polarization into higher-order polarization, which is applicable to obtaining arbitrary order polarization, and then, study the focusing properties of higher-order polarization under metasurface lens with numerical aperture (NA) from 0.1 to 0.9. 
It is shown that higher-order polarization generates smaller focusing spot than linear polarization (Table.~\ref{tab:comparsion}) due to the smaller radial intensity profile in the focal plane (Fig.~\ref{fig:higher_order_pol}d); more importantly, for $\mathrm{NA}<0.6$, the focusing spot of third- and fourth-order polarization can even break through the Abbe diffraction limit of the corresponding NA (Fig.~\ref{fig:higher_order_pol}d).
To further reduce the focusing spot of higher-order polarization, the effect of binary phase optical element added at the aperture of metasurface lens is studied as well, which enhances the imaging resolution strongly.
At last, the effect of substrate thickness and center deviation on the focusing feature of metasurface setup is investigated. 

\section{Theory}

\subsection{Transformation from linear to higher-order polarization}

\begin{figure*}[hbtp]
    \centering
    \includegraphics[width=0.75\linewidth]{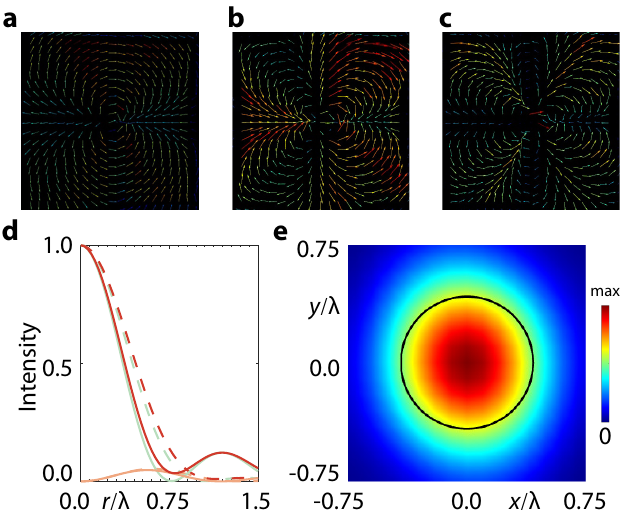}
    \caption{\textbf{Higher-order polarization and focusing properties.}
    \textbf{a-c.} Polarization distribution for (a) second-, (b) third-, and (c) forth-order polarized beam.
    \textbf{d.} Electric energy density profiles of forth-order (solid curves) and linearly (dashed curves) polarized beams at the focal plane of NA-0.6 lens, where light green curves denote the radial component $E_r^2$, brown curves for the longitudinal component $E_z^2$ and red curves for the total field $E_r^2+E_z^2$.
    \textbf{e.} Total electric energy distribution of fourth-order polarization at the focal plane of NA-0.6 lens, where the black circle indicates the Abbe limit at NA=0.6, indicating that the fourth-order polarization can reach Abbe diffraction limit at NA=0.6 without using any special technique.
    }
    \label{fig:higher_order_pol}
\end{figure*}

The electric field $\textbf{E}$ of a generic vectorial beam can be written as 
\begin{align}\label{eq:vectorE}
    \textbf{E}=\left[\begin{aligned}
        E_x\\
        E_y\\
        E_z
    \end{aligned}\right]=\left[\begin{aligned}
     &cos(p\phi+\phi_0) \\
     &sin(p\phi+\phi_0) \\
     &~~~~~~~~~0
    \end{aligned}\right],
\end{align}
where $p$ is the order of polarization, characterizing the number of cycles when polarization azimuth changes $2\pi$ along the circumference direction; and $\phi=\mathrm{atan}(y/x)$ is the azimuthal angle and $\phi_0$ is the initial azimuthal angle. 
The vectorial beam with $p>1$ is called higher-order polarized beam.
In this way, the classical radial and azimuthal polarizations are in fact the first-order polarization (\ie, $p=1$) with initial azimuthal angle $\phi_0=0$ and $\pi/2$, respectively. 
For better understanding, Figs.~\ref{fig:higher_order_pol}a-c depict the polarization distributions of $p=2~,3~,4$,respectively, with zero initial angle $\phi_0=0$.

Higher-order polarization can be generated in a variety of ways, for example, by converting circular polarization to higher-order polarization through a subwavelength grating~\cite{niv2003formation}, or by converting line polarization to higher-order polarization through a liquid crystal~\cite{davis2000two}.
We here report the generation of higher-order polarization from linear polarization incidence by means of metasurface~\cite{zhang2022chiral,xie2021polarization}. As shown below, the relation between polarization order $p$ and the rotation angle $\theta$ metasurface element acts to impose certain phase shifting is very simple, so this method applies to obtaining arbitrary order polarization. 

\begin{figure*}[bhtp]
    \centering
    \includegraphics[width=0.75\textwidth]{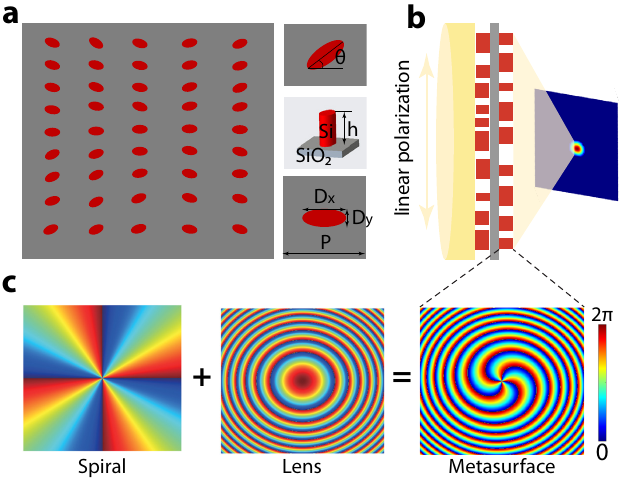}
    \caption{\textbf{Schematic of metasurfaces and imposed phase distributions.}
    \textbf{a.} Schematic of metasurface that transforms linear polarization into higher-order polarization. The metasurface is composed by elliptic silicon cylinders with same dimensions ($h=\SI{715}{nm}$, $D_x=\SI{124}{nm}$ and $D_y=\SI{83}{nm}$), but being rotated by different angle $\theta$ with respect to $x$ axis, and the elliptic cylinders are resided at the centers of square unit cells with period $\mathrm{P}=\SI{650}{nm}$. 
    \textbf{b.} Schematic of transforming and focusing linear polarization into high-resolution spot by metasurfaces.
    \textbf{c.} The formation of the focusing metasurface phase profile by superposing the phase profiles of spiral phase plate and focusing lens.
    }
    \label{fig:schematic_metasurface}
\end{figure*}

Using the notation of Jones matrix, the relation between incident and transmitted electric fields of metasurface reads 
\begin{align}\label{eq:EoutEin}
    \left[\begin{aligned}
        E_{out}^x\\
        E_{out}^y
    \end{aligned}\right]=T\left[\begin{aligned}
        E_{in}^x\\
        E_{in}^y
    \end{aligned}\right],
\end{align}
where $E_{in/out}^{x/y}$ is the $x/y$ component of input/output electric fields. For dielectric metasurface, the elements work as a linearly birefringent wave plate~\cite{mueller2017metasurface}, hence, can be described as follow
\begin{align}\label{eq:modumatrix}
    T=R(\theta)\left[\begin{array}{cc}
        e^{\iu\psi_x} & 0 \\
        0 & e^{\iu\psi_y}
    \end{array}\right]R(-\theta),
\end{align}
where $\psi_{x/y}$ denotes the propagation phase element imposes on light linearly polarized along its fast and slow axes that are rotated by an angle $\theta$ with respect to the reference coordinate (Fig.~\ref{fig:schematic_metasurface}a), and $R(\theta)$ is the two-dimensional rotation matrix.

In principle, a vectorial beam of any form, including higher-order polarization, can be derived from Eqs.~\ref{eq:EoutEin} and \ref{eq:modumatrix} as long as the information (\ie, $\psi_x$, $\psi_y$, $\theta$) of field modulation matrix $T$ are given.
We here are devoted to only higher-order polarization.
To this aim, the metasurface element plays the role of half-wave plate, \ie, $\psi_x-\psi_y=\pi$, the modulation matrix gets to
\begin{align}\label{eq:Tmattix}
    T=\iu\left[\begin{array}{cc}
        cos(2\theta) & sin(2\theta) \\
        sin(2\theta) & -cos(2\theta)
    \end{array}\right].
\end{align}
Substitute above equation into Eq.~\ref{eq:EoutEin} and consider the incidence of $x$-axis linear polarization, we can get the transmitted electric field
\begin{align}\label{eq:Eout}
    E_{out}=e^{\iu\pi/2}\left[\begin{aligned}
        cos(2\theta)\\
        sin(2\theta)
    \end{aligned}\right].
\end{align}
By comparing Eqs.~\ref{eq:vectorE} and \ref{eq:Eout}, we can see that Eq.~\ref{eq:Eout} is nothing but higher-order polarized beam with zero initial angle $\phi_0=0$. 
Thus, when transmitted through the metasurface consisting of half-wave plates, $x$-polarized beam is transformed into higher-order polarization, whose polarization order can be determined by
\begin{equation}\label{eq:polorder}
    p=(2\theta-\phi_0)/\phi.
\end{equation}
Note that both $\theta$ and $\phi$ are functions of coordinates $x$ and $y$. Accordingly, by means of Eq.~\ref{eq:polorder} and the desired polarization order $p$, it is convenient to determine the rotation angle $\theta$ of each element, and in turn the metasurface that transforms linear polarization into higher-order polarization. 

\subsection{Focusing of higher-order polarization}
One of the attractive advantages of vectorial beam is high-resolution imaging~\cite{liu2022super}. 
When tight focusing under high-NA lens occurs, vectorial diffraction theory~\cite{richards1959electromagnetic} that considers the vectorial feature of electric and magnetic fields should be used. Based on this theory, the electric field of higher-order polarization in the focal plane reads
\begin{align}\label{eq:focusingelec}
    \textbf{E}(r,\phi,z)=\frac{-\iu A}{\pi}\int_0^{\alpha}E_0(\frac{f \mathrm{sin}\theta}{\omega_0})^{2n+|m|}\mathrm{exp}(-\frac{f^2 \mathrm{sin}^2\theta}{\omega_0^2})\mathrm{sin}\theta\mathrm{cos}^{\frac{1}{2}}\theta Q_{p,m}(r,\phi,\theta)e^{-\iu kz \mathrm{cos}\theta}d\theta,
\end{align}
where $A$ is a constant, $m$ represents the topological charge and $n$ is the beam order, $\omega_0$ is the beam waist ($m=n=0$) of Gaussian beam, $r$ and $\phi$ are the radial and angular coordinates of cylindrical reference respectively, and $Q_{p,m}(r,\phi,\theta)$ is the polarization matrix~\cite{xu2022tight}. 

Shown in Fig.~\ref{fig:higher_order_pol}d are the electric energy density profiles of fourth-order (solid curves) and linearly (dashed curves) polarized beams at the focal plane of NA-0.6 lens.
The FWHM (full width at half maximum) of total electric field of fourth-order polarization (solid red curve in Fig.~\ref{fig:higher_order_pol}d) is about 0.84$\lambda$ ($\lambda=\SI{915}{nm}$ is the wavelength of incident light), which is smaller than that of linear polarization (FWHM= 0.93$\lambda$), even approximating the Abbe limit of NA-0.6 lens (0.84$\lambda$, black circle in Fig.~\ref{fig:higher_order_pol}e). 
We would like to stress that the feature of smaller focusing spot of fourth-order polarization is also true, compared to circular~\cite{stafeev2022flat}, radial~\cite{chen2017tight} and azimuthal~\cite{cheng2015tight} polarization.

In addition, from Fig.~\ref{fig:higher_order_pol}d we see that for both fourth-order and linear polarizations, the radial component (green curves in Fig.~\ref{fig:higher_order_pol}d) dominates the focusing field; nevertheless, the longitudinal component (brown curves in Fig.~\ref{fig:higher_order_pol}d) prevents the electric field from generating smaller spot; this is quite different from the focusing feature of radially polarized Bessel-Gaussian beam~\cite{wang2008creation}, in which the longitudinal component dominates the focusing field and enables the forming of smaller spot.

\section{Results and discussion}
\subsection{conversion of higher-order polarization from linear polarization by metasurface }
Now we numerically demonstrate the transformation of linear polarization into higher-order polarization by means of dielectric metasurface.
From Eqs.~\ref{eq:Eout} and \ref{eq:polorder} we see the critical point of realizing polarization transformation is modulating the phase profile of the incidence, which is implemented through the array of element of metasurface.
As shown in Fig.~\ref{fig:schematic_metasurface}a, the metasurface is composed of identical elliptic silicon cylinders that are mounted at silica substrate in square-lattice pattern. 
To implement the conversion of polarization, each cylinder first needs to work as a half-wave plate, \ie, $\psi_x-\psi_y=\pi$ (Eq.~\ref{eq:polorder}).
For this aim, the cylinders are all shaped into elliptic cross section with major and minor semi-axes, say, being $D_x=\SI{124}{nm}$ and $D_y=\SI{83}{nm}$ (bottom inset of Fig.~\ref{fig:schematic_metasurface}a),  respectively. 
In addition, the cylinders should be locally rotated with respect to $x$ axis by an angle $\theta(x,y)$ (top inset of Fig.~\ref{fig:schematic_metasurface}a) that can be determined from the polarization order $p$ and initial angle $\phi_0$ of the desired higher-order polarization (Eq.~\ref{eq:polorder}) to modulate the phase profile of incident beam. 
Based on the dielectric metasurface constructed above, higher-order polarization with arbitrary order can be produced with the incidence of linear polarization, as exemplified by Fig.~\ref{fig:higher_order_pol}a-c

\subsection{Tight focusing of higher-order polarization}
As aforementioned, the higher-order polarization is characterized with smaller focusing spot, compared to those of other polarizations, say, linear~\cite{zhuang2020tight}, circular~\cite{stafeev2022flat}, radial~\cite{chen2017tight} and azimuthal~\cite{cheng2015tight} polarization.
Before proceeding, we would like to point out that the higher-order polarization generated through Eqs.~\ref{eq:Eout} and \ref{eq:polorder} is associated with a helical phase structure that results from the Pancharatnam-Berry phase~\cite{berry1984quantal}, which alters the beam center during propagation, so needs to be cancelled out~\cite{niv2003formation}. 
Thus, besides the phase profile of focusing lens, the metasurface lens also needs to present the phase of spiral plate. 
In momentum space, the phase of electromagnetic field satisfies the principle of linear superposition~\cite{xie2021polarization}, so the phase profile of focusing metasurface lens is just the addition of the phase profiles of spiral phase plate~\cite{zhang2018generating} and focusing lens~\cite{aieta2012aberration}, \ie, 
\begin{equation}\label{eq:metaphase}
    \Phi(x,y)=m\cdot \mathrm{arctan}(\frac{y}{x})+k_0(\sqrt{x^2+y^2+f^2}-f),
\end{equation}
where $(x,y)$ is the coordinate of the cylinder center with respect to the origin locating at the center of metasurface, $m$ is the topological charge of spiral plate, $k_0=2\pi/\lambda$, and $f$ is the focal length of lens.
Figure~\ref{fig:schematic_metasurface}c shows the phase profile of metasurface lens with $m=4$ and $f=\SI{32}{\mu m}$.

Based on the dielectric metasurface lens constructed above, we compute the focusing of higher-order polarization with $p=2, 3, 4, 5, 6$ by metasurface lens of distinct NA, and the results are presented in table~\ref{tab:comparsion}.
For better comparison, the focusing of linear polarization and Abbe diffraction limit of distinct NA are also displayed in the second and last row, respectively. 
It is clear that higher-order polarization gives rise to smaller focusing spot than linear polarization due to the fact that higher-order polarization generates smaller radial component (Fig.~\ref{fig:higher_order_pol}d). 

Another remarkable feature in table~\ref{tab:comparsion} is that for $\mathrm{NA}\leq0.6$, the focusing spot of third- and fourth-order polarization is smaller that Abbe limit of the corresponding NA. 
This feature can find many applications for higher-order polarization. For instance, by replacing linear polarization with third or fourth-order polarization as source, microscope can exhibit higher resolution under the same NA, but with little cost~\cite{liu2022super}~\cite{chen2013imaging}.

\begin{table}
    \centering
    \caption{FWHM of focusing spot from metasurface under distinct incident polarization (unit: $\lambda$).}
    \begin{tabular}{p{2cm}|p{1cm}|p{1cm}|p{1cm}|p{1cm}|p{1cm}|p{1cm}|p{1cm}|p{1cm}|p{1cm}}
    \hline
         NA & 0.1 & 0.2 & 0.3 & 0.4 & 0.5 & 0.6 & 0.7 & 0.8 & 0.9\\
         \hline
         linear & 4.61 & 2.59 & 1.74 & 1.33 & 1.08 & 0.93 & 0.83 & 0.8 & 0.74\\
         \hline
         2nd order  & 4.59 & 2.50 & 1.63 & 1.28 & 1.09 & 0.93 & 0.85 & 0.77 & 0.72\\
         \hline
         3rd order & 4.48 & 2.20 & 1.46 & 1.15 & 0.95 & 0.86 & 0.78 & 0.75 & 0.66\\
         \hline
         4th order & 4.50 & 2.23 & 1.49 & 1.17 & 0.95 & 0.84 & 0.76 & 0.73 & 0.63\\
         \hline
         5th order  & 4.61 & 2.32 & 1.52 & 1.18 & 0.98 & 0.87 & 0.78 & 0.74 & 0.65\\
         \hline
         6th order & 4.55 & 2.31 & 1.54 & 1.20 & 0.99 & 0.85 & 0.79 & 0.74 & 0.65\\
         \hline
         Abbe limit & 5.00 & 2.50 & 1.67 & 1.25 & 1.00 & 0.83 & 0.71 & 0.63 & 0.56\\
         \hline
    \end{tabular}
    \label{tab:comparsion}
\end{table}

\subsection{Enhancement of resolution using binary-phase modulation}
In many optical applications, smaller focusing spot is always desired~\cite{pahlevaninezhad2018nano}~\cite{wang2014orbit}~\cite{chang2014ultra}. 
From Fig.~\ref{fig:higher_order_pol}d, we see that the radial electric component dominates the focusing field, so reducing the FWHM of radial component will get smaller spot. 
We here show that this is possible by using an additional binary optical element at the lens aperture~\cite{wang2008creation}.

\begin{figure*}
    \centering
    \includegraphics[width=0.8\linewidth]{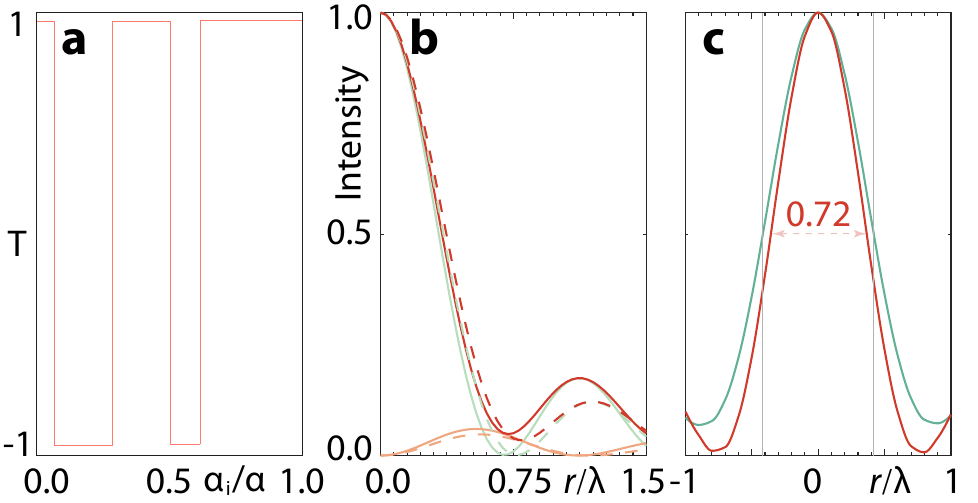}
    \caption{\textbf{Focusing properties of fourth-order polarization by NA-0.6 lens with using binary-phase modulation.}
    \textbf{a.} Amplitude transmission function $T$ of binary-phase modulation plate as a function of relative focusing angle $\alpha_i/\alpha$, where $\alpha$ is the largest aperture angle of lens.
    \textbf{b.} Electric energy density profiles of fourth-order polarized beam at the focal plane of NA-0.6 lens, where solid curves are for using phase modulation and dashed curves without using phase modulation. Light green curves denote the radial component $E_r^2$, brown curves for the longitudinal component $E_z^2$ and red curves for the total field $E_r^2+E_z^2$. 
    \textbf{c.}  Total electric energy density profiles of fourth-order polarized beam focused by NA-0.6 lens with (red) and without (green) phase modulation, and the vertical light grey lines indicate the range of Abbe diffraction limit, where the full width at half maximum is $0.72\lambda$, breaking Abbe limit by 13.3$\%$ or so.
    }
    \label{fig:fwhm_binary_phase}
\end{figure*}

To add binary-phase modulation, the aperture apodization function is imposed an additional transmission function $\mathrm{T}(\beta)$, which takes 1 or -1 for different ranges of angle $\beta$.
We use a five-belt optical element with transmission function $\mathrm{T}$ satisfying.
\begin{equation}\label{eq:transT}
    \mathrm{T}(\beta)=\begin{cases}
        1,&{\text{for}} ~{0\leq\beta<\alpha_1,\alpha_2\leq\beta<\alpha_3,\alpha_4\leq\beta<\alpha},\\
       -1,&{\text{for}}~{\alpha_1\leq\beta<\alpha_2,\alpha_3\leq\beta<\alpha_4}.
    \end{cases}
\end{equation}
As shown in Fig.~\ref{fig:fwhm_binary_phase}a, the four angles $\alpha_i~(i=1,...,4)$ correspond to four radial positions $r_i=sin\alpha_i/\mathrm{NA}$ (normalized to the optical aperture). 
By optimizing these four angles, the ratio of the radial and longitudinal components can be further enhanced to get smaller focusing spot.
For instance, Fig.~\ref{fig:fwhm_binary_phase}b shows the electric energy intensity after the modulation of binary-phase plate with $\alpha_1=\SI{2.80}{\degree}$, $\alpha_2=\SI{6.46}{\degree}$, $\alpha_3=\SI{11.02}{\degree}$, $\alpha_4=\SI{24.92}{\degree}$. 
Note that the optimization is performed with respect to NA=0.6, whose largest aperture angle $\alpha=\SI{36.87}{\degree}$. 
The FWHM of radial component energy profile in Fig.~\ref{fig:fwhm_binary_phase}b is 0.65$\lambda$, which is smaller than 0.71$\lambda$ shown in Fig.~\ref{fig:higher_order_pol}d. 
The FWHM of the total electric energy profile is 0.72$\lambda$, which is smaller than that obtained without using binary-phase modulation (0.84$\lambda$), being reduced by 14.3$\%$.

\begin{figure}[htbp]
    \centering
    \includegraphics[width=0.65\linewidth]{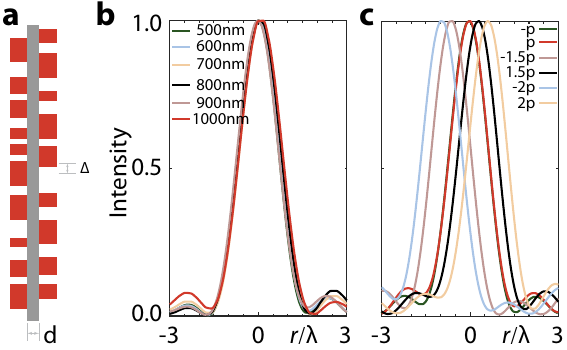}
    \caption{\textbf{Effect of substrate thickness and central deviation of transforming and focusing metasurfaces on the focusing property.}
    \textbf{a.} Schematic of bilayer metasurfaces, where the front one transforms incident linear polarization into higher-order polarization and the rear one focuses higher-order polarization. $d$ and $\Delta$ denote the thickness of substrate and central deviation of these two metasurface.
    \textbf{b.} Total electric energy density of third-order polarization focused by NA=0.3 metasurface lens for distinct thickness $d$, which implies that the FWHM of third-order polarization exhibits strong thickness tolerance.
    \textbf{c.} Total electric energy density of third-order polarization focused by NA=0.3 metasurface lens for distinct central deviation $\Delta$. While the spot center alters with $\Delta$, the alteration of FWHM for different $\Delta$ is less than 2$\%$ And $\mathrm{P}$ denotes lattice period of metasurface.
    }
    \label{fig:focus_defects}
\end{figure}

\subsection{Effect of substrate thickness and central deviation}
In optical application, the metasurface elements are normally with the dimension of subwavelength, hence, the metasurfaces proposed above can not always be fabricated perfectly. In this section we consider the effect of fabrication accuracy on the focusing features. 

In general, besides the element size difference between the ideal and fabricated cylinders, there are two kinds of factor that may affect the focusing spot drastically,\ie, the substrate thickness $d$ and the central deviation of these two metasurface layers $\Delta$, as shown in Fig.~\ref{fig:focus_defects}a.
As aforementioned, the element size difference is already included in the local phase difference between theoretical and realistic phases of elements ($\pm\SI{3}{\degree}$) when constructing metasurfaces, so we consider the later two factors only.

To this aim, we construct several metasurfaces with distinct substrate thicknesses and central deviations, and simulate the focusing of third-order polarization under NA-0.3 metasurface lens to investigate their focusing spot variation as the variation of $d$ and $\Delta$ (in unit of $p$).
The results are shown in Figs.~\ref{fig:focus_defects}b and c.
We see that for different thickness $d$, the FWHM of focusing spot remains the same (Fig.~\ref{fig:focus_defects}b), implying that the proposed metasurface setup exhibits strong thickness tolerance. 
While the focusing spot shifts for distinct central deviation $\Delta$ (Fig.~\ref{fig:focus_defects}c), the FWHM does not change as $\Delta$ (FWHM variation is less than $2\%$). 
In applications, the spot shiftness can be compensated by slightly moving the sample to be studied, so the resolution of optical system using higher-order polarization as source will not change since the FWHM of spot not alter.

\section{Conclusion}
In summary, we derived the formula of transforming linear polarization into higher-order polarization, which is very simple and applicable to generating arbitrary-order polarization, and studied the focusing features of distinct higher-order polarizations based on metasurface lens of different NA.
It is shown that compared to linear, circular, radial and azimuthal polarization, higher-order polarization gives rise to smaller focusing spot, and even breaks Abbe diffraction limit for NA$<$0.6.
By adding a binary-phase plate on the metasurface lens aperture, the focusing spot of higher-order polarization can be further reduced, say, by 14.3$\%$ for fourth-order polarization.
In addition, the effect of substrate thickness and central deviation of metasurface setup on the focusing property of higher-order polarization is also investigated, and the results shows that the focusing of higher-order polarization exhibits strong thickness tolerance.
Our proposed metasurface with large tolerance of substrate thicknesses (d) and central deviations ($\Delta$) may find potential applications in high-resolution lithography and imaging, say, by replacing the conventional linear or circular polarization source with higher-order polarization. 

\bibliographystyle{apsrev4-2}
\bibliography{Reference}
\end{document}